\documentclass{article}
\usepackage{graphicx,emulateapj}

\def\lxt{$L_{X}-T_{X}$}
\def\micmT{$M_{ICM}-T_{X}$}
\def\rfive{$r_{500}$}
\def\TX{$T_{X}$}
\def\fICM{$f_{ICM}$}

\submitted{Submitted April 7, 2000; Accepted June 28, 2000; to appear 
in ApJ Dec. 1, 2000}
\lefthead{MOHR, REESE, ELLINGSON, LEWIS \& EVRARD}
\righthead{X-RAY ST RELATION AT INTERMEDIATE REDSHIFT}
\def\myputfigure#1#2#3#4#5%
{\hskip0.03\textwidth\vskip#5pt
\makebox[0pt]{\hskip#2in
\includegraphics[width=#3\textwidth]{#1}}\vskip#4pt\hfill}

\begin{document}

\title{The X-ray Size-Temperature Relation for Intermediate Redshift Galaxy 
Clusters}

\author{Joseph J. Mohr\altaffilmark{1,2}, Erik D. Reese\altaffilmark{2}, 
E. Ellingson\altaffilmark{3}, Aaron D. Lewis\altaffilmark{3} \& \\
August E. Evrard\altaffilmark{4}}

\altaffiltext{1}{Chandra Fellow}
\altaffiltext{2}{Department of Astronomy and Astrophysics, University of 
Chicago, Chicago, IL}
\altaffiltext{3}{Department of Astronomy, University of Colorado, Boulder, CO}
\altaffiltext{4}{Department of Physics, University of Michigan, Ann Arbor, MI}

\authoremail{mohr@oddjob.uchicago.edu}
\authoremail{reese@oddjob.uchicago.edu}
\authoremail{e.elling@casa.colorado.edu}
\authoremail{lewisad@colorado.edu}
\authoremail{evrard@umich.edu}

\begin{abstract}
We present the first measurements of the X-ray size-temperature (ST)
relation in intermediate redshift ($z\sim0.30$) galaxy clusters.  
We interpret the local ST relation ($z\sim0.06$) in terms of 
underlying scaling relations in the cluster dark matter 
properties, and then we use standard models for the 
redshift evolution of those dark matter properties to show that the 
ST relation does not evolve with redshift.  We then use ROSAT HRI 
observations of 11 clusters to examine the intermediate redshift 
ST relation; for currently favored cosmological parameters, the intermediate 
redshift ST relation is consistent with that of local clusters.   
Finally, we use the ST relation and our evolution model to measure angular diameter 
distances; with these 11 distances we evaluate constraints on 
$\Omega_{M}$ and $\Omega_{\Lambda}$ which are consistent with
those derived from studies of Type Ia supernovae.
The data rule out a model with $\Omega_{M}=1$ and $\Omega_{\Lambda}=0$ with 
2.5$\sigma$ confidence.  When limited to models where 
$\Omega_{M}+\Omega_{\Lambda}=1$, these data are inconsistent with 
$\Omega_{M}=1$ with 3$\sigma$ confidence. 
\end{abstract}


\section{Introduction}

Nearby galaxy clusters exhibit a tight correlation between X-ray 
isophotal size and emission weighted intracluster medium (ICM) temperature 
(\cite{mohr97}; hereafter ME97).  This 
correlation is evidence of regularity; it
exists in an X-ray flux 
limited sample of 45 clusters (\cite{edge90}), where no attempt has 
been made to use the X-ray morphologies to exclude
clusters showing signatures of recent, major mergers.  The scatter 
around the X-ray size--temperature (ST) relation is approximately 
15\% in size, comparable to 
the scatter of elliptical and lenticular galaxies around their fundamental plane 
(\cite{jorgensen96}).  This small scatter in the galaxy cluster 
scaling relation is intriguing, because (1) there is overwhelming 
evidence that galaxy clusters are still accreting mass 
(e.g. \cite{mohr95,buote96}) and (2) 
elliptical galaxies are generally thought to be among the most 
regular objects in the universe.

ME97 use 48 N-body and hydrodynamical simulations of 
cluster formation in 
four different cosmological models to address this apparent 
contradiction between regularity and ongoing accretion in nearby 
clusters.  Using simulations from both 
$\Omega_{M}=0.3$ and $\Omega_{M}=1$ cosmologies, they show that a tight ST 
relation is expected even in cosmogonies where 
there is significant cluster growth at the present epoch.

The high degree of regularity implied by the ST relation is surprising, because 
the well known correlation between X-ray luminosity and emission 
weighted mean temperature (the \lxt\ relation) has very large scatter 
(\cite{david93}).
ME97 show that the same cluster ensemble which exhibits a ~15\% scatter
in the ST relation exhibits a 52\% scatter in $L_{X}$ around 
the \lxt\ relation.  This higher 
scatter in the \lxt\ relation results from the sensitivity of the 
X-ray luminosity to the densest regions of the cluster-- a 
sensitivity to the presence or absence of so-called cooling flows 
(\cite{fabian94}).  This interpretation is supported by more recent 
work where cluster ensembles specially chosen to contain no cooling 
flow clusters conform to \lxt\ relations with 
significantly reduced scatter of ~25\% (\cite{arnaud99}).  
Additionally, when central parts of cooling 
flow clusters are excluded, the scatter in the \lxt\ relation 
decreases (\cite{markevitch98}). 

Finally, cluster regularity is also 
evident in the tight correlation between ICM mass and 
emission weighted temperature (the \micmT\ relation).  When measuring 
ICM mass within a limiting radius of \rfive\ (the radius where the 
enclosed overdensity is 500 times the critical density) the scatter 
in mass about the \micmT\ relation is 17\% (\cite{mohr99}).
Observational studies of the ST relation followed by work on the 
\lxt\ and \micmT\ relations support a scenario where 
clusters exhibit regularity similar to that of elliptical galaxies 
in the properties measured on the scales of their virial regions, but 
exhibit significant irregularities in the properties of the densest, central 
regions where physical processes other than gravity and gas 
dynamics-- such as radiative cooling and magnetic fields-- play 
significant roles (\cite{mohr97,arnaud99,mohr99}).  
The evidence indicating cluster regularity 
is balanced by evidence for departures from regularity; the scatter in 
the observed scaling relations is larger than can be accounted 
for by the measurement uncertainties.  This resolved scatter 
contains clues about, among other things, cluster peculiar velocities 
and departures from equilibrium.

Here we examine the ST relation at intermediate redshift ($0.19\le 
z\le0.55$) using ROSAT 
HRI observations of the Canadian Network for Observational Cosmology  
(CNOC; e.g. \cite{yee96,lewis99}) cluster sample.
The ST relation provides a potentially powerful tool to study the 
expansion history of the universe.  As explained in detail below, the 
ST relation is rather insensitive to cosmological evolution.
Thus, armed with an accurate model of cluster evolution, one could 
use the ST relation to measure distances at intermediate redshift, 
constraining the deceleration parameter $q_{0}$.

We first present the X-ray ST relation in the nearby cluster sample 
($\S$\ref{sec:localST}) and then present an interpretation of the ST 
relation in terms of regularity in the underlying dark matter 
properties of the cluster.  Section \ref{sec:midzST} describes observations of 
the ST relation in intermediate redshift clusters.  
In $\S$\ref{sec:cosmo} we use these 
observations to constrain cosmological parameters.  
Section \ref{sec:conclude}  contains a summary of our 
conclusions.  Throughout the paper we use 
$H_{0}=50h_{50}\,\rm{km/s/Mpc}$.

\myputfigure{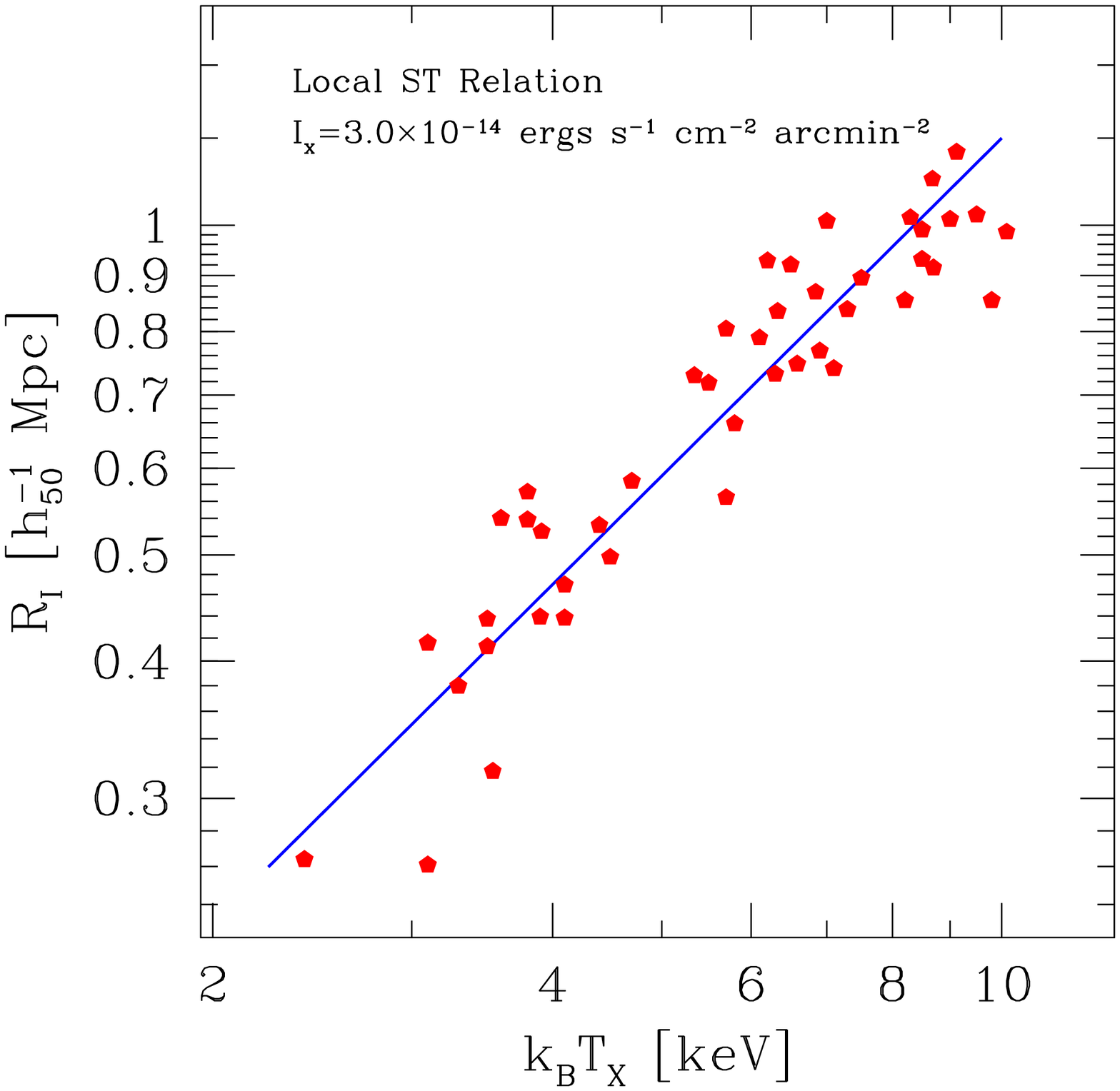}{2.8}{0.45}{-20}{-10}
\figcaption{The X-ray isophotal size versus emission weighted mean 
ICM temperature $T_{X}$ for an X-ray flux limited sample of 45 nearby 
clusters.  We use the isophote 
$I=3.0\times10^{-14}$~erg/s/cm$^{2}$/arcmin$^{2}$ within the cluster 
rest frame 0.5:2.0~keV band.  The solid line represents the best fit 
relation, and the RMS scatter about this line is 15\% in size.
\label{fig:STnearby}}

\section{Local X-ray ST Relation}
\label{sec:localST}
Below we present observations of the ST relation in an X-ray flux 
limited sample of clusters and then discuss
cluster scaling relations.  

\subsection{Observations}
Fig \ref{fig:STnearby} contains a plot of the X-ray ST relation for 
an ensemble of nearby galaxy clusters (see ME97 for cluster list).  
This sample is essentially a low redshift sample, 
$z_{median}\sim\left<z\right>=0.055$ with RMS variation $\sigma_{z}=0.03$, 
but the full redshift range is $0.01\le z\le0.19$.
The cluster isophotal size $R_{I}$ is plotted versus the emission weighted mean 
ICM temperature \TX, where
\begin{equation}
R_{I}=\sqrt{A_{I}/\pi}
\label{eq:size}
\end{equation}
and $A_{I}$ is the area enclosed by the isophote $I$.  The isophotal 
size is a model-independent measure extracted directly from the X-ray 
image. ICM temperatures come from the literature (listed in ME97).  
In this figure 
we use the galactic absorption corrected surface brightness
$I=3.0\times10^{-14}\,\rm{ergs}~\rm{s}^{-1}\rm{cm}^{-2}\rm{arcmin}^{-2}$ within the 
cluster rest frame 0.5-2.0~keV band.   In the conversion 
from count rate to physical flux 
units we use PROS, assume a Raymond-Smith spectrum with the published 
mean temperature $T_{X}$ and ${1\over3}$ cosmic abundances 
(\cite{mush97}).  We convert between the angular size and physical 
size of the isophotal region using the angular diameter 
distance, and we also correct $I$ for cosmological dimming $(1+z)^{4}$.

The best fit power law to this local ST relation has the form
\begin{equation}
R_{I}=(0.71\pm0.02)\left({T_{X}\over{6\,\rm keV}}
\right)^{1.02\pm 0.11}h_{50}^{-1}{\rm \ Mpc}
\end{equation}
and the scatter about this relation is 15\% in $R_{I}$.  This fit is 
shown as a solid line in Fig \ref{fig:STnearby}.  
Because this is the local ST relation, it is only 
weakly dependent on the deceleration parameter.
For this relation we 
have assumed the parameters $\Omega_{M}=0.3$ and 
$\Omega_{\Lambda}=0.7$;  for $\Omega_{M}=0.3$ and $\Omega_{\Lambda}=0$
the normalization is 
1.8\% lower, and for $\Omega_{M}=1$ the normalization is 2.7\% lower.

\subsection{Cluster Scaling Relations}
\label{sec:scaling}
Galaxy cluster scaling relations between, for example, virial mass 
and temperature $T$ are expected if clusters are 
approximately virialized.  The spherical collapse model predicts that 
newly collapsed objects will have mean densities which are $\Delta$ times 
the critical density $\rho_{crit}=3H^{2}/8\pi G$, where $H$ is the 
Hubble parameter and in an Einstein-de~Sitter 
model $\Delta=18\pi^{2}$.  Therefore, we define the virial radius 
$r_{\Delta}$ of an object with virial mass $M_{\Delta}$ to be
\begin{equation}
r_{\Delta}=\left({M_{\Delta}\over{4\over3}\pi\Delta\rho_{crit}}\right)^{1/3}.
\label{eq:rdelta}
\end{equation}
Newly collapsed clusters may also satisfy the virial relation
\begin{equation}
 GM_{\Delta}=aTr_{\Delta}
 \label{eq:virial}
\end{equation}
where $T$ is the virial temperature and $a$ is a number dependent on
the cluster density and temperature structure; in the case that 
cluster structure is self similar (i.e. that massive clusters are 
simply rescaled versions of low mass clusters) $a$ will be constant for all $T$ 
and Eqns \ref{eq:rdelta} and \ref{eq:virial} lead to the well known 
scaling relations between virial mass, radius and temperature
\begin{equation}
M_{\Delta}={\left(a'T\right)^{3\over2}\over 
\sqrt{\Delta\rho_{crit}}}\ {\rm and}\ 
r_{\Delta}={\left(a'T\right)^{1\over2}\over\sqrt{\Delta\rho_{crit}}},
\label{eq:scaling} 
\end{equation}
where $a'$ is a dimensionless structure parameter like $a$.  These scaling relations 
are expected to hold for a range of $r_{\delta}$ and $M_{\delta}$, 
where $r_{\delta}$ is the region with enclosed overdensity $\delta$  with 
respect to $\rho_{crit}$ and $M_{\delta}$ is the 
corresponding enclosed mass.
Numerical cluster simulations indicate that the emission 
weighted mean ICM temperature \TX\ is an adequate proxy for the virial 
temperature $T$, and that these scaling relations exist even in cluster populations
which exhibit merging and substructure (\cite{evrard96,bryan98}).  
Using simulations to normalize these relations, one finds that clusters 
with \TX\ of 10~keV have virial masses 
$M_{200}=4\times10^{15}h^{-1}_{50}M_{\odot}$ and virial radii 
$r_{200}=4h^{-1}_{50}Mpc$.
The small scatter in the observed ST relation, the correlation 
between ICM mass and temperature (see Section 1), and a direct study 
of the correlation between binding mass (estimated using cluster galaxy 
kinematics) and emission weighted temperature (\cite{horner99}) provide 
evidence that the conditions required for Eqn \ref{eq:scaling} to be valid 
are generally met in real galaxy clusters.

\subsection{X-ray ST Relation}
\label{sec:STscaling}
The X-ray surface brightness (units: ergs~s$^{-1}$cm$^{-2}$sr$^{-1}$) 
at a projected radius $R_{\delta}$ from the 
cluster center is
\begin{equation}
I(R_{\delta})={1\over2\pi\left(1+z\right)^{4}}\int_{0}^{\infty}\,dl\, 
n_{e}(r)n_{H}(r)\Lambda(T)
\end{equation} 
where $n_{e}$ and $n_{H}$ are the electron and proton number 
densities, $r=\sqrt{R_{\delta}^{2}+l^{2}}$ is the distance from the 
cluster center and $\Lambda(T)$ is the X-ray emission coefficient which 
includes contributions from thermal bremsstrahlung and line emission 
(\cite{raymond77}).  
We can express the ICM density $\rho_{g}$ in terms of 
the underlying dark matter density $\rho_{dm}$:  
$\rho_{g}(r)=f_{g}\rho_{dm}(r)g(r)$,
where $f_{g}$ is the ICM mass fraction within the virial region 
and $g(r)$ is a function 
describing differences in the dark matter and ICM density 
distributions; then, using $n_{i}=\rho_{g}/\mu_{i}m_{p}$, we can 
rewrite the X-ray surface brightness as
\begin{equation}
I(R_{\delta})={f_{g}^{2}\Lambda(T_{X})\over2\pi m_{p}^{2}\mu_{e}\mu_{H}
\left(1+z\right)^{4}}\int_{0}^{\infty}\,dl\, \rho^{2}_{dm}(r) g^{2}(r)
\end{equation}
We bring the emission coefficient out of the integral and use the value 
at $T=T_{X}$; although this is only strictly valid for an isothermal gas, the 
temperature insensitivity of $\Lambda(T)$ band limited to 0.5:2.0~keV 
makes this an excellent approximation even in the presence of 
departures from isothermality (e.g. \cite{fabricant80,mohr99}).

Finally, we express the dark matter density profile in terms of the 
characteristic overdensity of the virial region: 
$\rho_{dm}(r)=\Delta\rho_{crit}f_{dm}(r_{\Delta}y)$, where 
$f_{dm}(y)$ encodes the dependence of the dark matter profile on a 
dimensionless radius $y=r/r_{\Delta}$;  this approach is consistent 
with numerical cluster simulations, which indicate that cluster dark 
matter density profiles have a ``universal'' form (\cite{navarro97}, hereafter 
NFW).  More recently, higher resolution simulations have shown differences 
between the inner profile (at radii which are 1\% of the virial 
radius) and the form proposed by NFW (\cite{moore99,jing99}); these 
differences appear to be smallest for cluster scale haloes, and none 
of our conclusions are sensitive to the behavior of the 
density profile at 1\% of the virial radius.  Expressing $\rho_{dm}$ in this 
way, and removing the characteristic scale $R_{\delta}$ from the 
integral, the X-ray surface brightness becomes
\begin{equation}
I(R_{\delta})={f_{g}^{2}\Lambda(T)\rho_{crit}^{2}\Delta^{2}R_{\delta}
\over2\pi m_{p}^{2}\mu_{e}\mu_{H}\left(1+z\right)^{4}}\Theta
\label{eq:idelta}
\end{equation}
where $\Theta$ is a dimensionless integral which encodes the 
shape of the ICM density profile: 
$\Theta=\int_{0}^{\infty}\,d\lambda\, f_{dm}^{2}(\eta) g^{2}(\eta)$, where
$\eta=\sqrt{1+\lambda^{2}}$ and $\lambda^{2}=\left(r/R_{\delta}\right)^{2}-1$.

We relate $R_{\delta}$ and $R_{I}$ using the shape of the 
typical X-ray surface brightness profile; $I(R)$ is typically well fit by the 
so-called $\beta$ model (\cite{cavaliere78})
\begin{equation}
I(R)=I_{0}\left(1+\left({R\over R_{c}}\right)^{2}\right)^{-3\beta+1/2}
\label{eq:beta}
\end{equation}
where $R_{c}$ is the core radius;  well outside the core the 
surface brightness falls as $I(R)\propto R^{1-6\beta}$.  Therefore we 
write
\begin{equation}
R_{I}=R_{\delta}\left(I(R_{\delta})\over I\right)^{1/(6\beta-1)}.
\end{equation}
Using Eqn.~\ref{eq:scaling} to substitute \TX\ for $R_{\delta}$, one 
can readily determine the ST relation at a particular redshift
in the case that cluster structure is self similar
\begin{equation}
R_{I}\propto T_{X}^{\alpha}\,{\rm where }\, \alpha={3\beta\over6\beta-1}
\end{equation}
For $\beta=2/3$, a typical observed value (\cite{jones84,mohr99}), the 
predicted ST relation has $\alpha=2/3$, 
shallower than the observed relation: $\alpha=1.02\pm0.11$.  
The observed increase in ICM mass fraction $f_{g}$
with cluster temperature \TX\ (\cite{mohr99}) is enough 
to explain the steepness of the observed relation.  The physics 
underlying this systematic variation in cluster gas mass fraction is 
likely heating of the intergalactic medium during the process of 
galaxy formation.  This so-called preheating affects the structure of 
the gas in low mass clusters more so than in high mass clusters 
(e.g. \cite{metzler94,cavaliere98,ponman99}).  As discussed below, in 
comparing the local and intermediate redshift ST relations it is 
important only that the effects of this preheating be similar in both 
samples.

\section{Intermediate Redshift X-ray ST Relation}
\label{sec:midzST}
Before presenting the intermediate redshift ST relation, we present
an evolution model and our measurement methods.

\subsection{Evolution of the ST Relation}
\label{sec:STevolve}
Comparison of ST relations in nearby and distant populations requires 
a model for how the expected changes in cluster structure with 
increasing redshift will affect the ST relation.  In the 
previous section we presented an explanation of the correlation 
between cluster X--ray isophotal size and emission weighted ICM 
temperature \TX\ as a manifestation of underlying scaling relations in 
the dark matter properties of the cluster.  Following this line of 
reasoning, we now express the evolution of the ST relation in terms of 
the redshift evolution of these same dark matter properties.  

The evolution of the dark matter scaling relations is apparent in 
Eqn \ref{eq:scaling}.  The normalization of these scaling relations 
changes with redshift according to the change in $\rho_{crit}$ and the 
change in the characteristic overdensity $\Delta$ of collapsed 
haloes.  Within Einstein-de~Sitter models $\Delta$ does not evolve, 
whereas evolution is expected in low $\Omega_{M}$ models.  
For simplicity of presentation, in the following analysis we 
explicitly follow only changes in $\rho_{crit}$, but accounting for 
changes in $\Delta$ would not affect our conclusions.  
Using $H(z)=H_{0}E(z)$, we express 
the evolution of the mass and size scaling relations as
\begin{equation}
M_{\Delta}(T,z)={M_{\Delta}(T,0)\over E(z)}\,{\rm and }\,
R_{\Delta}(T,z)={R_{\Delta}(T,0)\over E(z)}
\label{eq:scaleevolution}
\end{equation}
where 
$E^{2}(z)=\Omega_{M}(1+z)^{3}+
(1-\Omega_{M}-\Omega_{\Lambda})(1+z)^{2}+\Omega_{\Lambda}$, and 
$\Omega_{M}$ and $\Omega_{\Lambda}$ are the present epoch contributions 
to the density parameter from matter and ``dark energy'', 
respectively.  These simple evolution models are valid only for self 
similar evolution-- i.e. in the case that distant cluster dark matter 
profiles are structurally similar to those of nearby clusters.

\myputfigure{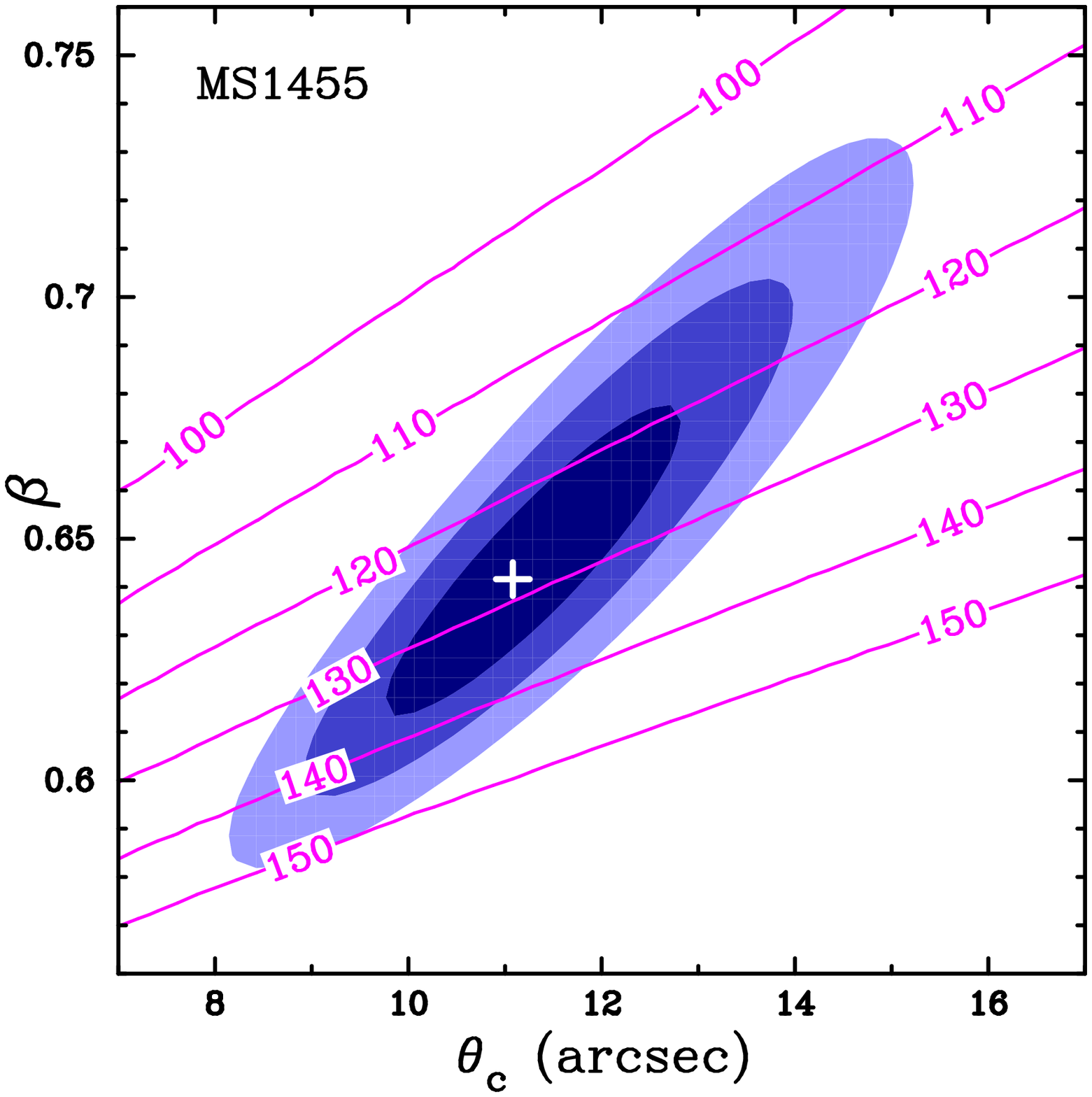}{2.8}{0.45}{-10}{0}
\figcaption{The 1, 2 and 3 $\sigma$ joint confidence regions for the 
$\beta$ model fit to MS-1455 ($\Delta\chi^{2}=2.3$, 6.2, 11.8).  
Overplotted are lines of constant 
isophotal size $\theta_{I}$ in arcseconds.  Note that these lines run at shallow 
angles with respect to the $\beta$-$\theta_{c}$ correlation, 
reducing the correlation's effect on our $\theta_{I}$ uncertainties.
\label{fig:uncertain}\vskip5pt}

From Eqn \ref{eq:idelta} it follows that the X-ray surface brightness 
at $R_{\delta}$ evolves as $I(R_{\delta})\propto E^{3}(z)$ and that the 
normalization of the ST relation evolves as 
\begin{equation}
R_{I}(T,z)=R_{I}(T,0)\,E^{\eta}(z)\, {\rm where }\, 
\eta={4-6\beta\over 6\beta-1}
\end{equation}
Interestingly, for the most common cluster profile $\beta=2/3$ 
(local: \cite{jones84,mohr99}; this intermediate redshift sample: 
$\left<\beta\right>=0.63\pm0.04$),
so $\eta=0$, corresponding to no evolution in the ST relation.  
In essence, a cluster of a given \TX\ is denser in the 
past, boosting its X-ray surface brightness and tending to increase 
the isophotal radius $R_{I}$, but this cluster is also smaller in the 
past, tending to decrease the isophotal radius $R_{I}$.  If the ICM 
density profile behaves as $\rho_{g}\propto r^{-2}$, 
these two effects cancel.  Thus, a population of clusters with 
measured $\beta\sim2/3$ will exhibit an ST relation which is 
relatively insensitive to cosmological evolution.  In principle, this 
behavior makes the ST relation ideal for measuring angular diameter 
distances to high redshift clusters.

\myputfigure{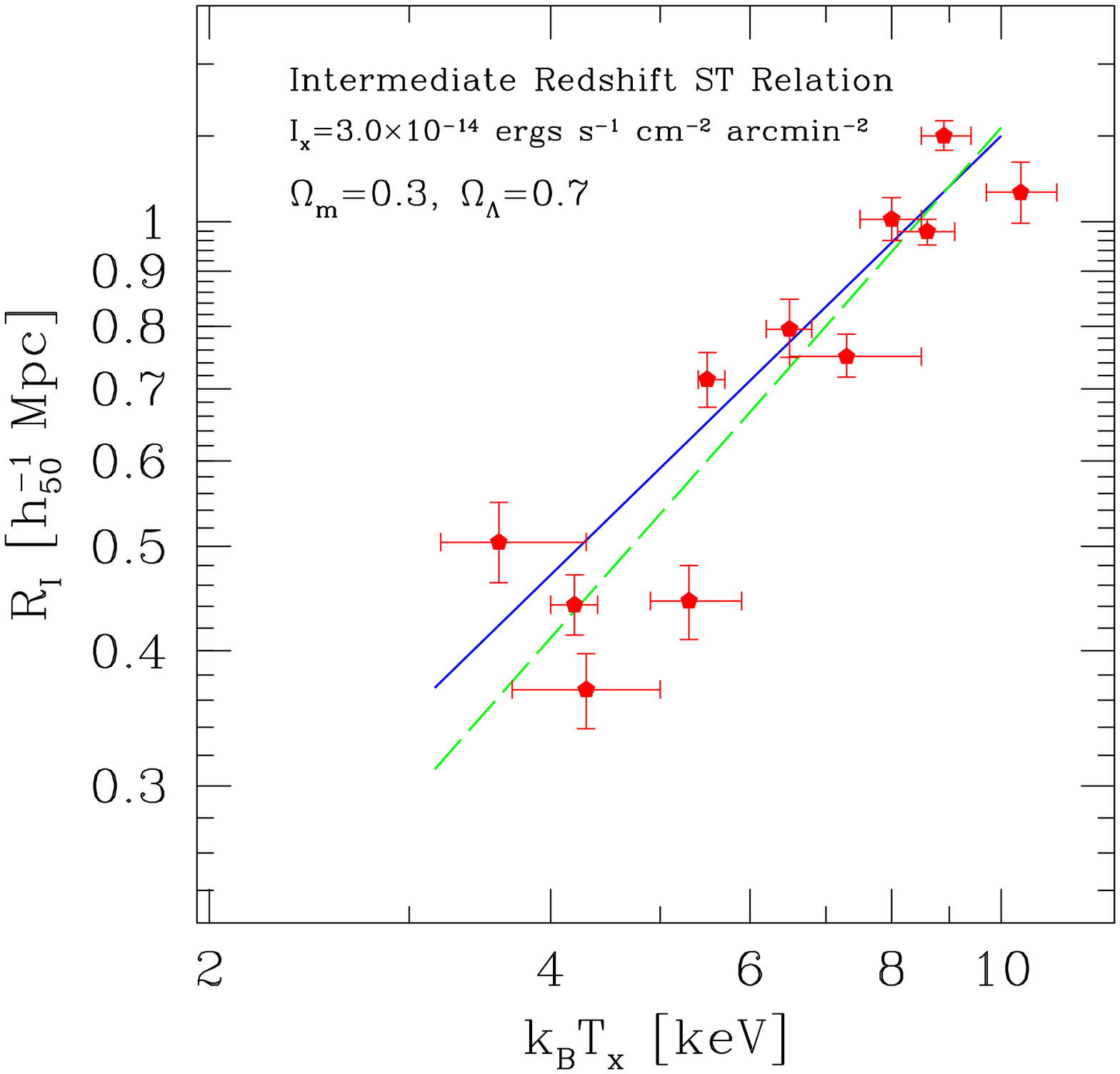}{2.8}{0.45}{-20}{-10}
\figcaption{The X-ray ST relation for 11 members of the CNOC cluster 
sample with measured \TX.  For this figure the conversion from measured 
$\theta_{I}\rightarrow R_{I}$ is done assuming $\Omega_{M}=0.3$ 
and $\Omega_{\Lambda}=0.7$.
Error bars denote 1 $\sigma$ uncertainties in emission weighted mean 
temperature $T_{X}$ and in isophotal size $R_{I}$.
The solid line is the best fit ST relation determined from the nearby cluster 
sample (Fig \ref{fig:STnearby}), and the dashed line is the best fit 
ST relation for the intermediate redshift sample.
\label{fig:STmidz}\vskip0pt}

\subsection{Observations}

We use archival ROSAT HRI observations of 11 of the 14 CNOC clusters discussed in 
Lewis et al. (1999); these 11 clusters are those with published \TX\ 
measurements. 
A detailed description of the reduction 
to 0.5$''$ pixel images is given in Lewis et al. (1999).  
The HRI angular resolution is $\sim5''$ FWHM; we further bin 
these images to a final pixel size of 2$''$ 
without significant loss of angular resolution.  Despite the far 
higher median exposure time (33~ks compared to 8~ks), the image 
quality of these intermediate redshift clusters is far lower than for 
our nearby sample (due to combination of lower observer frame 
surface brightness and higher instrument background).
The image quality is too poor to allow
isophotal sizes measurements directly from the images using 
the approach described in 
$\S$\ref{sec:localST}; 
therefore, we fit circular $\beta$ models (same form as Eqn 
\ref{eq:beta} with $\theta$ substituted for $R$)
to these images using the software developed for measuring SZE+X-ray distances 
(\cite{reese99}).  Essentially, we find the set of 
parameters $I_{0}$, $\theta_{c}$, $\beta$, cluster centroid 
$(\alpha,\delta)$ and local background $I_{bkg}$ 
which maximizes the likelihood of consistency 
between model and data.  In all these fits we fix the background to 
the value measured in an annulus extending from 4.5$'$ to 5$'$, a 
background dominated region with an essentially negligible contribution from the 
cluster.  Fits are performed to the central 
portion ($\theta\le4.2'$) of the image which includes the cluster and a local 
background region.  In Table~\ref{tab:fitparam} 
we list the best fit parameters with estimates of the statistical 
uncertainties.  The central surface brightness and measured 
background are both given in detector units of cts~s$^{-1}$arcmin$^{-2}$.

These cluster parameters are in reasonably good agreement with those 
presented in Lewis et al. (1999), with the exception of two clusters 
A2390 and MS1358.  The different best fit parameters in these two 
clusters stem from the treatment of emission excesses; we fit the 
surface brightness model to the entire dataset, whereas Lewis et al. 
(1999) exclude the central region in both these 
clusters.  This exclusion approach typically leads to larger core radii and 
correspondingly higher $\beta$'s than would fitting to the entire 
cluster as in our method.  

\myputfigure{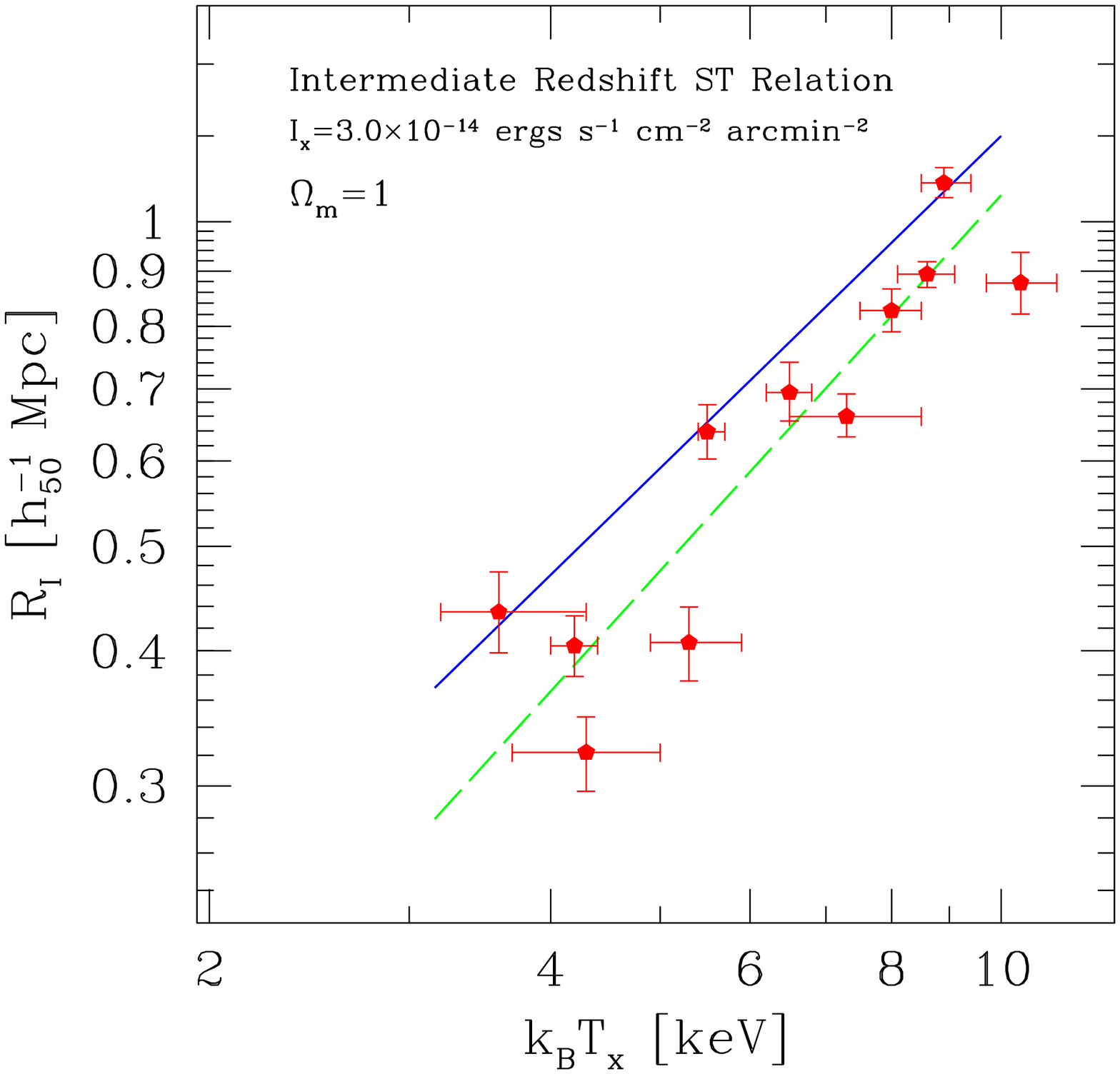}{2.8}{0.45}{-20}{-10}
\figcaption{The X-ray ST relation for 11 members of the CNOC cluster 
sample.  For this figure the conversion from measured 
$\theta_{I}\rightarrow R_{I}$ is done assuming $\Omega_{M}=1$.
Error bars denote 1 $\sigma$ uncertainties in emission weighted mean 
temperature $T_{X}$ and in isophotal size $R_{I}$.
The solid line is the best fit ST relation determined from the nearby cluster 
sample (Fig \ref{fig:STnearby}), and the dashed line is the best fit 
ST relation for the 11 intermediate redshift clusters.
\label{fig:STmidzSCDM}\vskip5pt}

An examination of the fit residuals in our sample 
provides no clear evidence for systematic differences between the data 
and the models with the exception of A2390, which exhibits an asymmetry 
or centroid variation (e.g. \cite{mohr93}).  
Allowing for an ellipticity introduces two additional free 
parameters, and the information content of some of the images simply 
is not sufficient to provide meaningful constraints.  Our goal here is 
to determine whether there is any evidence of regularity in intermediate 
redshift clusters;
any mismatch between the data and the model (like that in A2390) 
will serve simply as an additional source of scatter in the final relation.  
Therefore, we conservatively present a uniform and objective analysis which is 
appropriate for the majority of our sample, rather than varying 
our analysis from cluster to cluster. (An elliptical $\beta$ model 
fit to A2390 yields an isophotal size which is 8\% smaller than that 
listed in Table~\ref{tab:fitparam}.  This correction makes A2390 more 
consistent with the best fit ST relation, but the correction is 
unimportant compared to the 17\% RMS scatter of the sample about 
that best fit relation.)  Of course the best approach is 
a non-parametric analysis of the surface brightness maps such as that 
applied to the local sample, but this is not possible with the current 
data.

We use PROS to convert the detector count rate into
galactic absorption corrected flux units of ergs~s$^{-1}$cm$^{-2}$ 
within the cluster rest frame 0.5:2.0~keV band.  As for the low 
redshift clusters,
we assume a Raymond-Smith emission spectrum with the 
measured mean temperature $T_{X}$ and ${1\over3}$ cosmic 
abundances (\cite{mush97}) at the appropriate redshift, and then we 
correct for the $(1+z)^{4}$ cosmological dimming.  The rest frame 
0.5:2.0~keV central surface brightnesses in units of
ergs~s$^{-1}$cm$^{-2}$arcmin$^{-1}$ are also listed for each 
cluster.

\begin{deluxetable}{lccccccccr}
\tablewidth{0pt}
\tablecaption{Cluster Fit Parameters and Isophotal Size}
\tablehead{
 &
 &
\colhead{$k_{B}T_{X}$} &
 &
 &
\colhead{$\theta_{c}$} &
 &
 &
\colhead{$\theta_{I}$} &
\cr
\colhead{Cluster} &
\colhead{$z$}     &
\colhead{keV}     &
\colhead{$I_{o}$\tablenotemark{a}} &
\colhead{$I_{o}$\tablenotemark{b}} &
\colhead{$''$}    &
\colhead{$\beta$} & 
\colhead{$I_{bkgd}$\tablenotemark{a}} &
\colhead{$''$} &
\colhead{$R_{I}$\tablenotemark{c}}}
\footnotesize
\startdata
A2390         & 0.2279 &  \phn8.9$^{+0.5}_{-0.4}$ & 
0.159$^{+0.011}_{-0.008}$ &  \phn9.33 & 17.2$^{+1.1}_{-1.2}$ 
& 0.532$^{+0.008}_{-0.008}$ & 0.0022 & 235.1$^{+6.1}_{-5.7}$ & 1.202 \nl
MS0015.9+1609 & 0.5466 &  \phn8.0$^{+0.5}_{-0.5}$ & 
0.018$^{+0.001}_{-0.001}$ & \phn1.63 & 49.4$^{+5.9}_{-5.2}$ 
& 0.898$^{+0.095}_{-0.073}$ & 0.0020 & 112.3$^{+3.9}_{-3.6}$ & 1.005 \nl
MS0440.5+0204 & 0.1965 &  \phn5.3$^{+0.6}_{-0.4}$ & 
0.079$^{+0.020}_{-0.014}$ & \phn4.71 &  \phn9.1$^{+2.0}_{-1.7}$
& 0.521$^{+0.029}_{-0.023}$ & 0.0023 & \phn97.7$^{+5.8}_{-5.9}$ & 0.445 \nl
MS0451.5+0250 & 0.2010 &  \phn8.6$^{+0.5}_{-0.5}$ & 
0.012$^{+0.001}_{-0.001}$ & \phn0.69 & 94.6$^{+4.2}_{-4.2}$ 
& 0.750                     & 0.0024 & 211.0$^{+5.5}_{-5.8}$ & 0.979 \nl
MS0451.6-0305 & 0.5392 & 10.4$^{+0.8}_{-0.7}$ & 
0.025$^{+0.002}_{-0.002}$     & \phn2.30 & 36.4$^{+5.2}_{-4.1}$ 
& 0.752$^{+0.070}_{-0.056}$ & 0.0023 & 119.7$^{+5.7}_{-5.7}$ & 1.064 \nl
MS0839.8+2938 & 0.1928 &  \phn4.2$^{+0.2}_{-0.2}$ & 
0.129$^{+0.018}_{-0.015}$ & \phn6.33 & 11.9$^{+1.7}_{-1.4}$ 
& 0.588$^{+0.028}_{-0.022}$ & 0.0023 &  \phn98.4$^{+5.1}_{-4.9}$ & 0.442 \nl
MS1008.1-1224 & 0.3062 &  \phn7.3$^{+1.2}_{-0.8}$ & 
0.018$^{+0.001}_{-0.001}$ & \phn1.24 & 35.8$^{+4.1}_{-3.5}$ 
& 0.667$^{+0.042}_{-0.036}$ & 0.0023 & 118.5$^{+4.1}_{-3.7}$ & 0.745 \nl
MS1224.7+2007 & 0.3255 &  \phn4.3$^{+0.7}_{-0.6}$ & 
0.054$^{+0.013}_{-0.010}$ & \phn3.19 &  \phn7.5$^{+1.7}_{-1.3}$ 
& 0.552$^{+0.034}_{-0.028}$ & 0.0025 &  \phn55.9$^{+3.4}_{-3.6}$ & 0.368 \nl
MS1358.4+6245 & 0.3290 &  \phn6.5$^{+0.3}_{-0.3}$ & 
0.111$^{+0.022}_{-0.016}$ & \phn6.14 &  \phn8.4$^{+1.4}_{-1.4}$ 
& 0.501$^{+0.016}_{-0.016}$ & 0.0023 & 119.7$^{+6.0}_{-5.4}$ & 0.795 \nl
MS1455.0+2232 & 0.2570 &  \phn5.5$^{+0.1}_{-0.2}$ & 
0.460$^{+0.042}_{-0.037}$ & 31.86 & 11.2$^{+1.0}_{-0.9}$ 
& 0.643$^{+0.022}_{-0.019}$ & 0.0028 & 127.7$^{+6.3}_{-6.1}$ & 0.713 \nl
MS1512.4+3647 & 0.3726 &  \phn3.6$^{+0.7}_{-0.4}$ & 
0.107$^{+0.022}_{-0.017}$ & \phn6.20 &  \phn7.4$^{+1.4}_{-1.2}$ 
& 0.560$^{+0.032}_{-0.025}$ & 0.0022 &  \phn70.1$^{+4.7}_{-4.6}$ & 0.505 \nl
\enddata
\tablenotetext{a}{cts s$^{-1}$arcmin$^{-2}$}
\tablenotetext{b}{10$^{-12}$ ergs s$^{-1}$ cm$^{-2}$ arcmin$^{-2}$ in rest frame 
0.5:2~keV band}
\tablenotetext{c}{$h_{50}^{-1}$~Mpc for $\Omega_{M}=0.3$ and 
$\Omega_{\Lambda}=0.7$}
\label{tab:fitparam}
\end{deluxetable}

Angular isophotal sizes $\theta_{I}$ are then estimated using a 
fiducial surface brightness of 
$3.0\times10^{-14}$~ergs~s$^{-1}$cm$^{-2}$arcmin$^{-2}$.
We determine statistical uncertainties in the fit parameters by 
exploring the likelihood within a grid in the parameters 
$I_{0}$, $\theta_{c}$ and $\beta$; we then
use that range to estimate uncertainties in the 
derived isophotal size $\theta_{I}$.  Figure \ref{fig:uncertain} 
contains a plot of the 1, 2 and 3$\sigma$ confidence regions in 
$\beta$ and $\theta_{c}$ for MS1455; lines of 
constant isophotal size $\theta_{I}$ are overlaid.  Note that lines of constant 
$\theta_{I}$ are approximately parallel to the well known 
$\beta-\theta_{c}$ correlation (e.g. \cite{mohr99}), minimizing the 
detrimental effects this correlation 
has on our size measurement uncertainties. 

Measured $\theta_{I}$ are then converted into physical isophotal sizes 
$R_{I}$ using the angular diameter distance $d_{A}$:  
$R_{I}=\theta_{I}d_{A}$.  We use the general form
\begin{eqnarray}
d_{A}={c\over H_{0}(1+z)} {F\left(\kappa \int_{0}^{z} {dz'\over 
E(z')}\right)\over\kappa}\nonumber\\
\kappa=\sqrt{\left|1-\Omega_{M}-\Omega_{\Lambda}\right|}
\end{eqnarray}
where $F(x)=sinh(x)$, $x$, and $sin(x)$ in an open, flat and closed 
cosmology.
Table~\ref{tab:fitparam} contains a list of $\theta_{I}$ for 
$I_{x}=3.0\times10^{-14}$~ergs/s/cm$^{2}$/arcmin$^{2}$ with 
statistical uncertainty estimates and the corresponding $R_{I}$ for 
the case where $\Omega_{M}=0.3$ \& $\Omega_{\Lambda}=0.7$.

Figure \ref{fig:STmidz} contains the intermediate redshift ST relation
with 1$\sigma$ error bars in both $T_{X}$ and 
$R_{I}$.  The conversion from $\theta_{I}$ to $R_{I}$ assumes 
$\Omega_{M}=0.3$ and $\Omega_{\Lambda}=0.7$ in this figure.  The 
best fit relation in this case is
\begin{equation}
R_{I}=(0.67\pm0.04)\left({T_{X}\over{6\,\rm keV}}
\right)^{1.19\pm 0.21}h_{50}^{-1}{\rm \ Mpc},
\end{equation}
and the RMS scatter in size about this relation is 17\%.  Also 
plotted is the local ST relation for this same isophote (solid line).  The 
uncertainties are derived by bootstrap resampling the list of
11 sizes (allowing duplication) and refitting the relation.  

There is a 
suggestion that the intermediate redshift ST relation is steeper than the 
local relation, but the difference in slopes is less than 1$\sigma$ 
significant.  In fact, both the zeropoints and slopes of the local 
and intermediate ST relations are 
statistically consistent when we use this particular set of cosmological 
parameters.  
With a larger sample it will be possible to measure the 
intermediate redshift ST relation slope more accurately; a comparison of the two 
slopes would be an important diagnostic of unusual structural 
evolution in clusters.

Figure \ref{fig:STmidzSCDM} contains a plot of the intermediate 
redshift ST relation assuming $\Omega_{M}=1$.  The best fit slope in 
this case is $1.16\pm0.20$,  and the zeropoint is $0.59\pm0.034$.
This zeropoint differs from that of the local ST relation (also 
calculated with $\Omega_{M}=1$) at 2.5$\sigma$.

We have examined the local and intermediate redshift ST relations at 
four other isophotes: $1.5\times10^{-14}$~cgs, 
$2.0\times10^{-14}$~cgs, $4.0\times10^{-14}$~cgs, and $5.0\times10^{-14}$~cgs.
The general conclusions reached above using the isophote $3.0\times10^{-14}$~cgs
apply equally as well at these other isophotes:  there is a 
suggestion that the ST relation at intermediate redshift is steeper 
(at less than 1~$\sigma$ significance) than the local relation, and 
the zeropoints are in good agreement when using the cosmological 
parameters $\Omega_{M}=0.3$ and $\Omega_{\Lambda}=0.7$.  The fainter 
the isophote the larger the enclosed region.  Pushing to fainter 
isophotes is dangerous given the quality of the intermediate redshift 
data, and pushing to brighter isophotes leads to complications from
central emission excesses and the effects of the PSF.  Thus, we 
choose the isophote $3.0\times10^{-14}$~cgs as a compromise between 
these two competing effects.

\subsection{Cosmological Constraints}
\label{sec:cosmo}

Although the X-ray images of these 11 clusters are poor, it is 
nevertheless interesting to use the local ST relation and our 
evolution model to predict distances to the intermediate redshift 
sample, thereby constraining cosmological parameters.
Because the clusters are at a range of 
redshifts, both the scatter of the data about the local ST relation 
and any systematic offsets from the local relation contain 
cosmological information.  Operationally, for each set of cosmological parameters 
$\Omega_{M}$ and $\Omega_{\Lambda}$ we convert the measured 
$\theta_{I}\rightarrow R_{I}$ and calculate the $\chi^{2}$ of the 
sample about the best fit local ST relation, where the best fit local 
relation is calculated using this same set of cosmological parameters.   
Fig \ref{fig:cosmo} contains 
contours of $\Delta\chi^{2}$ corresponding to 1, 2 and 3 $\sigma$ 
confidence regions (equivalent to $\Delta\chi^{2}=2.3$, 6.2 and 11.8)
in the $\Omega_{M}$--$\Omega_{\Lambda}$ space.

The current sample, because of observational uncertainties, does not 
provide a strong cosmological constraint.  
The 1$\sigma$ confidence region is a wide trough similar 
in character to those derived from luminosity distances to SNe Ia 
(\cite{schmidt98,perlmutter99}), with the important difference that 
our sample of 11 intermediate redshift distances is not as 
constraining as the larger samples of SNe Ia distances.
Notably, there is enough power in the current dataset to
exclude the $\Omega_{M}=1$ \& $\Omega_{\Lambda}=0$ model (one of 
three models marked with stars) at $>$95\% confidence.
If one considers only models where $\Omega_{M}+\Omega_{\Lambda}=1$, 
then the preferred model has $\Omega_{M}=0.09$.  Furthermore, 
$\Omega_{M}<0.32$ with 1$\sigma$ confidence and $\Omega_{M}<1$ with 
3$\sigma$ confidence.

One can also use these data to constrain the deceleration parameter 
$q_{0}\equiv\Omega_{M}-\Omega_{\Lambda}/2$.  The preferred value is 
$q_{0}=0.11^{+0.41}_{-0.65}$ (68\% confidence).

Clearly, larger samples of archival data and the impending stream of 
high quality X-ray images from Chandra and XMM 
will provide tighter constraints.  One particular concern in this 
study is that the local and intermediate redshift samples were 
observed with two different instruments: the ROSAT PSPC and the ROSAT 
HRI.  Any relative calibration errors would then serve to bias the
preferred cosmological parameters.  However, the steepness of cluster 
X-ray surface brightness profiles mitigates this potential problem;  
typically, the surface brightness falls off as $I\propto\theta^{-3}$, 
so even a 10\% relative error in the calibration of the PSPC and HRI 
would introduce only a 3\% error in the angular size of the cluster.

Although there has been no previous use of cluster scaling relations 
to constrain cosmological parameters, there have been three attempts to use 
the ICM mass fraction \fICM\ (\cite{pen97,cooray98,rines99}); if one 
assumes \fICM\ is constant with redshift, then one can use its distance 
dependence to constrain the distance--redshift relation.
The Pen constraint $q_{0}=0.89\pm0.29$ is marginally inconsistent with our 
measurement, but 
the Pen analysis is only appropriate in the case that the ICM traces 
the dark matter, a case that is inconsistent with both 
theoretical and observational studies (\cite{david95,evrard97}).

In the Rines et al. analysis, 
care is taken to (1) match the luminosity ranges of the local and 
intermediate redshift samples, (2) address the expected radial 
variation of \fICM, and (3) test the consistency of the published 
effective areas of the two satellites involved (Einstein and ASCA).
Because \fICM\ increases slowly with radius, 
it is important that it be measured within the same portion of the 
virial region in the local and intermediate redshift samples; this 
complicates the use of \fICM\ measurements as cosmological constraints.
If one analyzes local and intermediate redshift clusters at a
fixed metric aperture, the region studied will 
correspond to a larger fraction of the virial region as the 
redshift increases (see Eqn.~\ref{eq:scaleevolution}).  
Similarly, when measuring \fICM\ within a fixed portion of the virial 
region \rfive, it is important that the redshift evolution of the
$r_{500}-T_{x}$ be included;
it appears that in both the Rines et al. and Cooray 
analyses, no evolution was included, so the
portion of the virial region used in 
determining \fICM\ increases with redshift.  In both cases, these 
biases cause the inferred angular diameter distances to be underestimated. 

The size of the bias depends on the cosmological
model being evaluated (Eqn.~\ref{eq:scaleevolution}); using the 
relation $f_{ICM}\propto d_{A}^{3/2}(r/r_{500})^{0.15}$ 
(\cite{evrard97,pen97}), we estimate the bias on 
$d_{A}$ at $z=0.35$ to be -2\% for $\Omega_{M}=0.3$ \& $\Omega_{\Lambda}=0.7$, 
and -10\% for $\Omega_{M}=1$ and $\Omega_{\Lambda}=0$; these biases 
are important relative to the $\sim$8\% difference in $d_{A}(z=0.35)$ 
between an $\Omega_{M}=0.3$ and an $\Omega_{M}=0.3$ \& 
$\Omega_{\Lambda}=0.7$ cosmology. 
Correcting for this bias will improve the consistency between our 
cosmological constraints and those of Rines et al. and Cooray.

\section{Conclusions}
\label{sec:conclude}

We present the local X-ray Size--Temperature (ST) relation for an 
X-ray flux limited sample of 45 clusters observed with the ROSAT 
PSPC (\cite{mohr97}).  We provide an explanation of this scaling 
relation in terms of underlying scaling relations in the cluster dark matter 
properties.  The observed ST relation is slightly 
steeper than the self--similar expectation presented in 
$\S$\ref{sec:STscaling}, but is consistent when the modest variation
of the ICM mass fraction  $f_{g}$ with $T_{x}$ is taken 
into account (\cite{mohr99}).

We use our theoretical model for the ST relation to explore its evolution with 
redshift.  Interestingly, for the typical ICM radial distribution 
observed in nearby and distant clusters, the normalization of the ST relation is 
not expected to evolve.  The lack of evolution makes the ST relation 
a plausible source of intermediate redshift angular diameter 
distances.  Of course, if cluster structure evolution is very different 
from the current theoretical expectation, our model 
($\S$\ref{sec:STevolve}) will underestimate ST relation evolution.
For example, a shift in the mean ICM mass fraction with redshift would
bias ST relation distances; to date there is no compelling evidence 
that distant clusters have different ICM mass fractions in the mean 
than nearby clusters (e.g. \cite{lewis99,grego00}). 

We use ROSAT HRI observations of 11 CNOC clusters with measured 
emission weighted mean ICM temperatures \TX\ to make the first 
measurements of the intermediate redshift ST relation.  Because of 
the poor image quality, we measure the angular isophotal size 
$\theta_{I}$ using the best fit circular $\beta$ model, rather than measuring 
it nonparametrically as for the nearby clusters (see Eqn~\ref{eq:size}).  
By assuming the cosmological parameters 
$\Omega_{M}=0.3$ \& $\Omega_{\Lambda}=0.7$, we
show that the slope and zeropoint of this intermediate redshift 
ST relation is statistically consistent with that of the local ST 
relation (see Fig~\ref{fig:STmidz}).  In addition, we examine the 
relation for $\Omega_{M}=1$ \& 
$\Omega_{\Lambda}=0$ (see Fig~\ref{fig:STmidzSCDM}), showing that 
although the slope is consistent, the zeropoint is different at 
greater than 3$\sigma$ significance.

\myputfigure{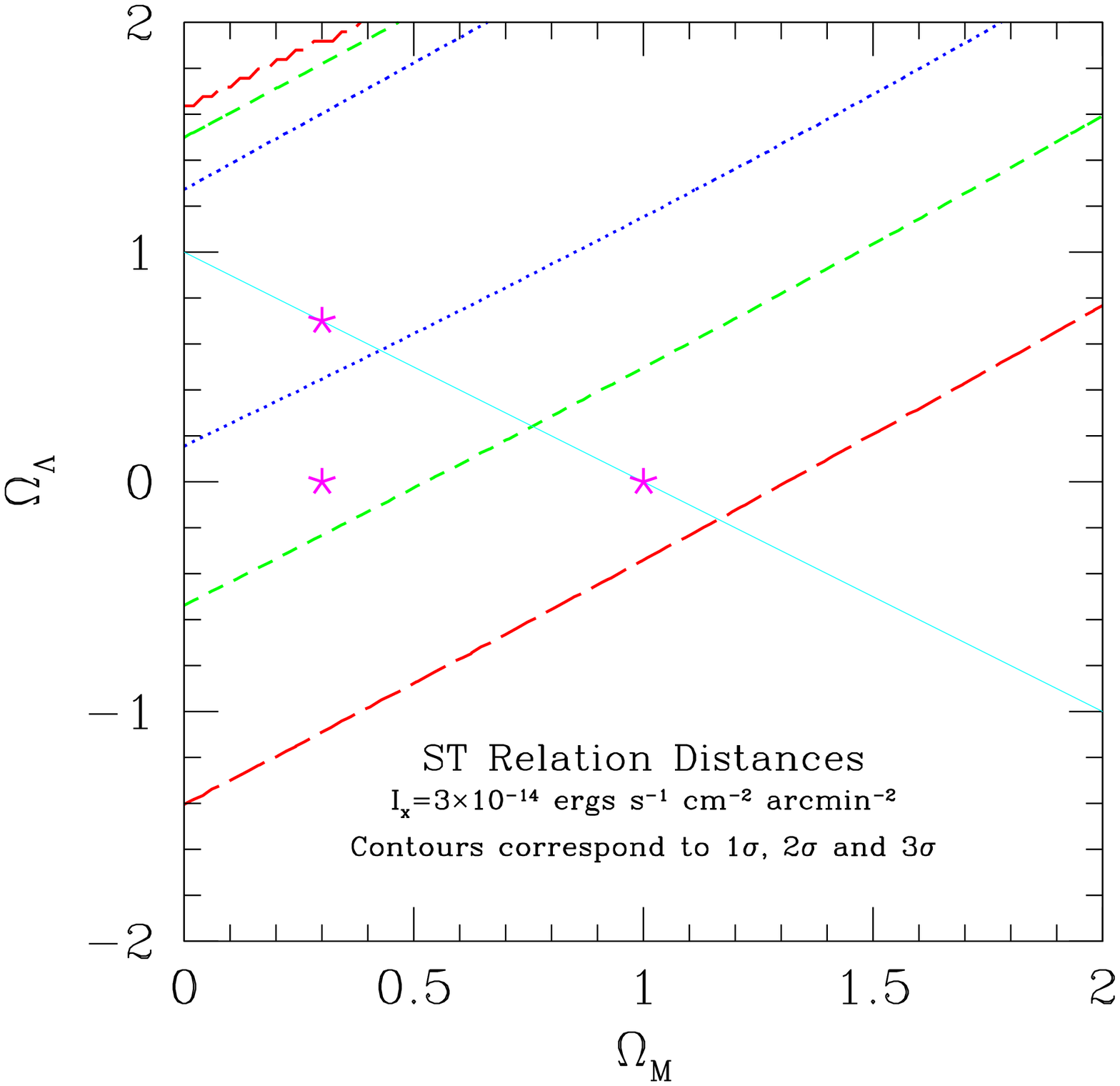}{2.8}{0.45}{-25}{-10}
\figcaption{Cosmological constraints from X-ray ST relation distance 
measurements to 11 intermediate redshift clusters.  Contours correspond to 
1, 2 and 3 $\sigma$ confidence regions ($\Delta\chi^{2}=2.3$, 6.2 and 
11.8).  The solid line marks spatially flat 
models, and the stars mark models $\Omega_{M}=0.3$ \& $\Omega_{\Lambda}=0.7$,
 $\Omega_{M}=0.3$ and  $\Omega_{M}=1$.
\label{fig:cosmo}\vskip5pt}

Finally, we use this cluster sample and our ST relation 
evolution model to place cosmological constraints.  Given the quality 
of the cluster images and temperature measurements, it is not surprising 
that a wide range of cosmological models is consistent with the data.
Nevertheless, this sample of 11 intermediate redshift distances is 
sufficient to rule out $\Omega_{M}=1$ with between 2$\sigma$ and 3$\sigma$ 
confidence.  Taken together with ICM mass fraction constraints on 
the cosmological matter density parameter $\Omega_{M}<0.44$ at 95\% confidence
(\cite{mohr99}), the cluster ST relation exhibits a slight
preference for universes with $\Omega_{\Lambda}>0$; models with 
$\Omega_{\Lambda}=0$ are inconsistent with the ST relation at 
between 1 and 2$\sigma$.  When considering only models where 
$\Omega_{M}+\Omega_{\Lambda}=1$, we can rule out $\Omega_{M}=1$ with 
3$\sigma$ confidence.

With the higher quality X-ray images and ICM temperature measurements 
available from Chandra and XMM, a significant tightening of these constraints 
and further tests of the underlying evolution model will be possible. 
Comparison of local and distant \micmT\ relations, which are more 
sensitive to cluster evolution, would provide important constraints on 
these models.   In addition, observations with a new generation of 
Sunyaev-Zel'dovich effect instruments 
(\cite{carlstrom99,mohr99b,holder00}) will allow us to 
more accurately constrain the evolution of cluster structure.
With these future observations of intermediate and high redshift 
clusters, we plan to continue using the ST relation as a tool to provide
cosmological constraints independent of those derived from recent high 
redshift SNe Ia observations (\cite{schmidt98,perlmutter99}).

\acknowledgements

JJM is supported by Chandra Fellowship grant PF8-1003, awarded 
through the Chandra Science Center.  The Chandra Science Center is 
operated by the Smithsonian Astrophysical Observatory for NASA under 
contract NAS8-39073.  EDR is supported by NASA GSRP Fellowship 
NGT5-50173.  EE acknowledges support provided by the National Science 
Foundation grant AST~9617145.  AEE acknowledges support from NSF 
AST-9803199 and NASA NAG5-8458.  This research has made use of data 
obtained through the High Energy Astrophysics Science Archive Research 
Center Online Service, provided by the NASA/Goddard Space Flight Center.

\end{document}